\documentclass[sn-mathphys-ay]{sn-jnl}

\usepackage{graphicx}%
\usepackage{multirow}%
\usepackage{amsmath,amssymb,amsfonts}%
\usepackage{bm}%
\usepackage{amsthm}%
\usepackage{mathrsfs}%
\usepackage[title]{appendix}%
\usepackage{xcolor}%
\usepackage{textcomp}%
\usepackage{manyfoot}%
\usepackage{booktabs}%
\usepackage{algorithm}%
\usepackage{algorithmicx}%
\usepackage{algpseudocode}%
\usepackage{listings}%

\graphicspath{ {./figures/} }
\DeclareMathOperator{\sech}{sech}

\theoremstyle{thmstyleone}%
\theoremstyle{thmstyletwo}%

\theoremstyle{thmstylethree}%

\begin{document}

\title{Model-based optimization of bacterial motility strategies for maximizing population yield}


\author[1]{\fnm{Peize} \sur{Yu}}\email{peize.yu.g7@elms.hokudai.ac.jp}

\author*[1]{\fnm{Sohei} \sur{Tasaki}}\email{tasaki@math.sci.hokudai.ac.jp}

\affil*[1]{\orgdiv{Department of Mathematics, Faculty of Science}, \orgname{Hokkaido University}, \orgaddress{\street{Kita 10, Nishi 8, Kita-ku}, \city{Sapporo}, \postcode{0600810}, \state{Hokkaido}, \country{Japan}}}


\abstract{
Bacterial motility is a fundamental trait for territorial expansion and resource acquisition. While existing models of nutrient-dependent motility often do not explicitly account for the metabolic costs associated with motility, these costs become critical in nutrient-limited or closed systems. In this study, we developed a mathematical framework using partial differential equations (PDEs) that explicitly incorporates the energetic trade-offs of motility. By formulating an optimization problem focused on maximizing population yield---defined by the total cell count---we evaluated various motility strategies across different environmental contexts. Our results demonstrate that the optimal motility response is highly sensitive to resource distribution. Specifically, we show that in unpredictable environments, a non-monotonic motility response emerges as the optimal strategy, providing a robust theoretical explanation for dose-response curves observed in experimental microbiology. This framework serves as a powerful, interpretable tool for predicting bacterial behavior in resource-constrained ecosystems and offers new insights into how such adaptive strategies are shaped by environmental pressures.
}




\maketitle

\section{Introduction}

Random (undirected) motility represents a fundamental strategy for bacteria to expand their colonial territory and explore environments with limited information, as exemplified by the formation of biofilms~\citep{tu2018adaptation}. A prominent example of regulated random motility is chemokinesis, a response where bacteria modulate their swimming speed based on the local concentration of chemoeffectors, typically nutrients~\citep{jakuszeit2021migration,ariel2015swarming}. While directed motility, such as chemotaxis, is extensively studied in the context of bacterial migration, recent evidence suggests that chemokinesis can significantly enhance sensing and navigation performance, particularly in environments characterized by spatial signal disturbance and noise~\citep{jakuszeit2021migration,hein2016physical}. Despite the rapid development of single-cell experimental techniques, the potential for macroscopic models of random motility to describe a broader range of behavioral scenarios remains largely unexplored~\citep{alt1980biased,sourjik2012responding,adler1966effect,adler1967effect,adler1969chemoreceptors}.

However, a motility response toward nutrients does not only offer benefits to a bacterial colony; it also entails significant metabolic costs. Recent research investigating bacterial motility in closed systems found inconsistencies between experimental results and predictions from modified PDE models that ignored the energetic investment in movement~\citep{tasaki2017self}. These findings indicate that the energy consumed by motility---including flagellar synthesis and operation---is a non-negligible factor, especially under nutrient-limited conditions. Consequently, motility can lead to resource inefficiency, potentially creating disadvantages for the colony depending on the environmental context.

Despite the scarcity of existing theoretical frameworks, there is a profound interest in the adaptive principles underlying these diverse motility patterns. The trade-off between motility intensity and metabolic efficiency is likely a key driver in the differentiation of behavioral strategies. By maximizing population-level advantages, such as population yield, bacteria may adopt specific strategies to survive under various selective pressures. Understanding how these strategic biases shape bacterial colonies can provide comprehensive insights into the fundamental principles of bacterial behavior.

In this work, we address three primary concerns identified in the current literature: (1) exploring the potential of a macroscopic framework based on modified random motility; (2) explicitly incorporating the metabolic consumption associated with motility into the model; and (3) interpreting the adaptive motivations behind different motility strategies in biased environmental settings. Inspired by classical capillary assay methods, we developed an adjusted PDE model that includes negative terms representing the energy consumption of motility within the nutrient equations~\citep{adler1967method,adler1973method,ordal1977chemotaxis,ordal1979chemotaxis}. By constructing a formal optimization problem, we aim to quantify and evaluate the advantages of various motility functions---which might otherwise remain ambiguous within an evolutionary scope---thereby providing new insights into the dynamics of bacteria and nutrients within resource-constrained systems.

\section{Development of Bacteria-nutrient Interaction Models with Nutrition-consuming Factors}

\subsection{Diffusion-growth-consumption Models}

In recent research studying how bacterial biofilm formation is influenced by environmental conditions~\citep{tasaki2017self}, a comprehensive mathematical model of bacterial motility under a condition with limited nutrients is applied: 
\begin{align}
&\frac{\partial B}{\partial t} + \nabla \cdot  (B {\bm u}) = P(B,N),     \\
&\frac{\partial N}{\partial t} = D_N \nabla^2N - k_pP(B,N), 
\end{align}
where $B=B(x,t)$ and $N=N(x,t)$ denote the bacterial concentration and the nutrient distribution, respectively, $x \in \Omega \subset {\mathbb R}^2$ is the space variable, $t \in {\mathbb R}$ is the time variable.   
In this model, $B$ is the bacteria move with the velocity ${\bm u} = {\bm u} (N, B)$ and proliferate consuming nutrition,
$P = P(B, N)$ represents the proliferation of bacteria,
nutrition $N$ diffuses in the media and is consumed by bacterial proliferation,
$D_N$ is the diffusion coefficient of nutrients,
$k_p$ represents the rate of nutrient consumption by proliferation.

The simulation results effectively mirrored the colony patterns and chemotactic behaviors seen in experiments with such a model, particularly the impact of chemotaxis on colony morphology under a wide range of environmental conditions~\citep{tasaki2017self}.
However, under environmental conditions with a certain level of acidity or higher, the simulation results no longer matched the experimental results. Observations suggest that, in the experiments, the colony failed to form a clear, crater-like image with the lack of population. After eliminating the other limiting factors, it was concluded that the discrepancy between simulation and experimental outcomes originates from the energy consumption of bacterial motility~\citep{tasaki2017self}. More specifically, this is due to the energy expended to produce ATP for the efflux of protons that enter the cell during flagellar movement, in order to maintain intracellular pH homeostasis. In this particular case, this factor will use up nutrients that are supposed to be used by bacteria for proliferation and maintenance of metabolism, distracting the mathematical model from simulating realistic results~\citep{tasaki2017self}.

\subsection{Nutrient-consuming motility model}

To represent this energy expenditure, we introduce a negative term into the nutrient equation of the classical PDE model, directly proportional to the motility intensity.
Reviewing existing models and clarify the hypothetical foundation for our comprehensive nutrient-consuming motility model, we have made following assumptions.

The spatio-temporal dynamics of the bacterial density, $B(x,t)$, and nutrient concentration, $N(x,t)$, are governed by a system of reaction-diffusion equations where local changes are driven by random motility, nutrient diffusion, and metabolic consumption.
To investigate the limits of minimal information processing, we assume bacteria lack the mechanism for gradient sensing (chemotaxis). Consequently, the motility intensity is a function of the local nutrient concentration, $N$, rather than its spatial gradient, $\nabla N$.
We restrict our analysis to undirected movement, which we approximate as a macroscopic diffusion process. Directed motility components, such as advection or biased runs, are neglected.
Nutrient depletion occurs through two distinct channels: the metabolic cost of cellular proliferation and the energetic expenditure required for flagellar-mediated motility.

These assumptions lead to the following coupled partial differential equations model:
\begin{align}
\frac{\partial B}{\partial t} &= \nabla \cdot ( D_B(B,N) \nabla B ) + P(B,N),     \\
\frac{\partial N}{\partial t} &= D_N \nabla^2 N - k_p P(B,N) - k_m D_B(B,N), 
\end{align}
where $D_B(B, N)$ represents the motility function, 
$P(B, N)$ represents the proliferation function, 
$D_N$ is the diffusion coefficient of nutrients, 
$k_m$ is the nutrient consumption coefficient of random motility,
$k_p$ is the nutrient consumption coefficient of proliferation.

\section{Optimization Problem Setup}

\subsection{General Problem}

The objective of this optimization framework is to determine the functional form of $D_B(B, N)$, which represents the bacterial motility strategy. We seek a strategy that maximizes the population-level benefits under given initial conditions within a specified timeframe. While several classical models treat $D_B(B, N)$ as a constant~\citep{lauffenburger1981effects,lauffenburger1982effects}, or represent strategies through the ratio of motile to immotile populations~\citep{mimura2000reaction}, our approach treats the motility response as a dynamic variable to be optimized against the associated metabolic costs.

In the following optimization variants, the adaptive strategy is defined by the coupling of the motility function $D_B(B, N)$ and the proliferation function $P(B, N)$. 
To quantify the success of a given strategy, we evaluate the population yield---defined as the total cell count---at three distinct time horizons: $t = T_{\rm Short}$, $T_{\rm Mid}$, and $T_{\rm Long}$. These represent short-, medium-, and long-term growth strategies, respectively, based on the assumption that a higher population yield enhances the collective survival and robustness of the colony. 

The performance metrics, or scores, for these time horizons are defined as: 
\begin{align}
S_J := B_{\rm total}(T_J) = \int_{\Omega} B(x,T_J) dx , 
\end{align}
where $J \in \left\{ {\rm Short},\  {\rm Mid}, \ {\rm Long} \right\}$. 
The analysis of these scores allows us to identify the theoretically optimal trade-off between motility intensity and metabolic efficiency. We hypothesize that the optimal strategy is highly sensitive to the evaluation period. For instance, in a closed system without resource replenishment, a longer time horizon may favor energy-saving strategies characterized by lower motility. Conversely, energy-consuming strategies may yield higher scores in the short term, as they prioritize rapid resource exploitation before nutrients are depleted.
To isolate the effects of motility on population dynamics, we assume a linear relationship between the proliferation rate and nutrient availability: 
\begin{eqnarray}
P(B, N) = \beta NB,
\end{eqnarray}
where $\beta$ is the coefficient controlling the proliferation rate. This simplification ensures that the proliferation response remains equally sensitive to nutrient levels across different scenarios, thereby reducing interference from irrelevant factors and allowing a focused evaluation of how various motility strategies shape the population yield.

\subsection{Linear Response to Nutrition}

To begin, we consider the framework of motility where its intensity is directly proportional to the amount of the nutrient~\citep{kawasaki1997modeling,rida2010effect,horger2015analysis}:  
\begin{align}
D_B(B,N) = \sigma BN . 
\end{align}
The mathematical model in this case is described as follows:
\begin{eqnarray}
\frac{\partial B}{\partial t} &=& \nabla \cdot (\sigma NB\nabla B) + \beta NB,     \\
\frac{\partial N}{\partial t} &=& D_N \nabla^2N - k_p \beta  NB - k_m \sigma NB , 
\end{eqnarray}
where $\sigma$ is the linear coefficient that controls motility according to the nutrient amount. 
Such a setting brings out the most fundamental motility strategy: the more nutrient supply available, the more energy to be used for motility. 
Although the linear response only represents a rather simple energy distribution, it suffices to be a good reference framework to evaluate other motility strategies and can be easily augmented for diverse purposes.

\subsection{Non-monotonic Response to Nutrition}

On the other hand, non-monotonic (also called as biphasic) responses are widely observed through different species and environments~\citep{mitchell2006bacterial}. Its inverted-U shape offers both quick reactions to initial stimulation from chemoattractors and suppressions at high concentration to stay at the most preferred position longer~\citep{yang2020biphasic,colin2021multiple}. To incorporate such motility feature, we reviewed the empirical data from existing research in~\cite{tasaki2017self}, and decide to use following model in which motility intensity is controlled by a hyperbolic secant function as it represents such a biphasic responses:

\begin{align}
D_B(B,N) = m \biggl(\frac{N}{\kappa+N^2}\biggr)^\alpha B . 
\end{align}
The model is described by
\begin{eqnarray}
\frac{\partial B}{\partial t} &=& \nabla \cdot \biggl\{m \biggl(\frac{N}{\kappa+N^2}\biggr)^\alpha B\nabla B\biggr\} + \beta NB,     \\
\frac{\partial N}{\partial t} &=& D_N \nabla^2N - k_p\beta NB - k_m\biggl\{m \biggl(\frac{N}{\kappa+N^2}\biggr)^\alpha B\biggr\} . 
\end{eqnarray}
In the model of the hyperbolic secant (biphasic) response, the motility coefficient is replaced by $m \biggl(\dfrac{N}{\kappa+N^2}\biggr)^\alpha B$. 
For $X = \log N$,
\begin{align*}
m \biggl(\frac{N}{\kappa+N^2}\biggr)^\alpha &= M\sech^{\alpha} (X-X_M);\\
M = m \biggl(\frac{1}{2\sqrt{\kappa}}\biggr), \quad X_M = \ln \sqrt{\kappa}, &\quad HWHM = \ln \biggl(2^{1/\alpha}+ \sqrt{4^{1/\alpha}-1}\biggr) , 
\end{align*}
where $M$ is the maximum value, $X_M$ the maximum point, and $HWHM$ the half width at half maximum.

The non-monotonic coefficient will demonstrate an inverted U-shaped curve on the $\log N$ axis, which describes a motility pattern that meets the prediction of a basic motility strategy:
When the nutrient is too low, motility will be suppressed to save enough energy for reproduction and maintenance.
When the nutrient is above a certain level, motility will be stimulated to expand territory and search for more resources.
When the nutrient is abundant enough in the logarithm scale, the marginal effect of motility to search and secure potential nutrients gradually reduces. A moderate level of suppression instead of a positive correlation will prevent meaningless motility and increase the general efficiency of energy consumption.

\section{Simulation and Results}

To assess the efficacy of various motility functions across diverse environmental configurations, we conducted numerical simulations in a one-dimensional spatial domain. The performance of each motility strategy was quantified using a metric defined as the population yield at specific time intervals:
\begin{equation}
 S_J = B_{\rm total} (T_J) = \int_{\Omega} B(x,T_J) dx, 
\end{equation}
where $B_{\rm total} (T_J)$ represents the total cell count at time $t=T_J$ ($J \in \left\{ {\rm Short},\  {\rm Mid}, \ {\rm Long} \right\}$).  
To capture the temporal dynamics of growth and resource consumption, we evaluated this score at three representative time points: short-term ($T_{\rm Short}= 7344$), mid-term ($T_{\rm Mid} = 11833$), and long-term ($T_{\rm Long} = 16285$). 
The simulations were implemented using the explicit Euler method. The following dimensionless parameters were employed to maintain consistency across all scenarios: 
$\Delta t = 0.001$, $N_x = 100$, $\Delta x = 0.1$, $\beta =1$, $D_N = 0.1$, $k_p = 1$, $k_m = 0.1$, $\alpha =1$.  
These settings allow for a rigorous comparison of how different motility response functions---specifically their sensitivity to nutrient gradients and their associated metabolic costs---influence the collective success of the bacterial colony.

\subsection{Linear Response to Nutrition}

A general search was first conducted to see the influence and tendency of changing in $\sigma$. We noticed that the curves of $S_J$ yield meaningful evidence of the general optimization condition.

\begin{figure*}[tbp]
\begin{center}
\includegraphics[width=135mm]{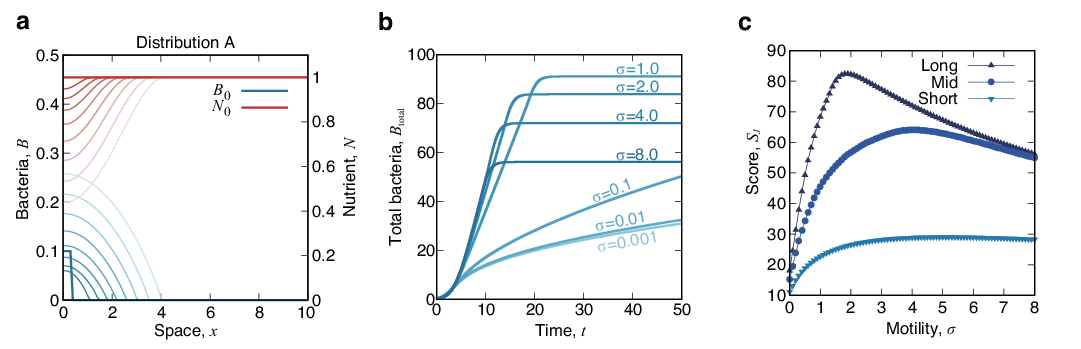}
\end{center}
\caption{Bacterial population model simulation of motility in linear response to nutrition. 
{\bf a}, Initial distribution of bacteria and nutrients, $B_0 = B_0 (x)$, $N_0 = N_0 (x)$. The thin curves with color gradations are the time-series data from the early stages. In particular, the initial nutrient distribution $N_0 = N_0 (X)$ is uniform here (distribution A). 
{\bf b}, Time series of total bacterial cell number for different motility. 
{\bf c}, Dependence of growth strategy score $S_J$ on motility, where $J \in \left\{ {\rm Short},\  {\rm Mid}, \ {\rm Long} \right\}$.}
\label{fig:1}
\end{figure*}

From the result of the general search, when $\sigma$ is small, the bacteria colony behaves passively towards nutrients, resulting in a slow but consistent growth curve. Population growth is mainly relying on the diffusion of nutrients, bringing distant nutrients slowly to the inoculation point. In an extremely safe condition where bacteria can survive long enough, low $\sigma$ preserves the most nutrients for proliferation.

As $\sigma$ increases, the slope of $B_{\rm total}$ starts to split into two parts, a rapidly increasing part for abundant nutrients ($N \gg 0$), and a long platform where $B_{\rm total}$ stops changing as $N\sim0$.

Under the limited time $t=50000$, $\sigma = 0.5$ ends up with the highest $B_{\rm total}$. Moreover, as $\sigma$ increases from 0.5, each curve intersects the previous one from the left, having a higher $B_{\rm total}$ before the intersection $t$ and vice versa.

\begin{figure*}[tbp]
\begin{center}
\includegraphics[width=135mm]{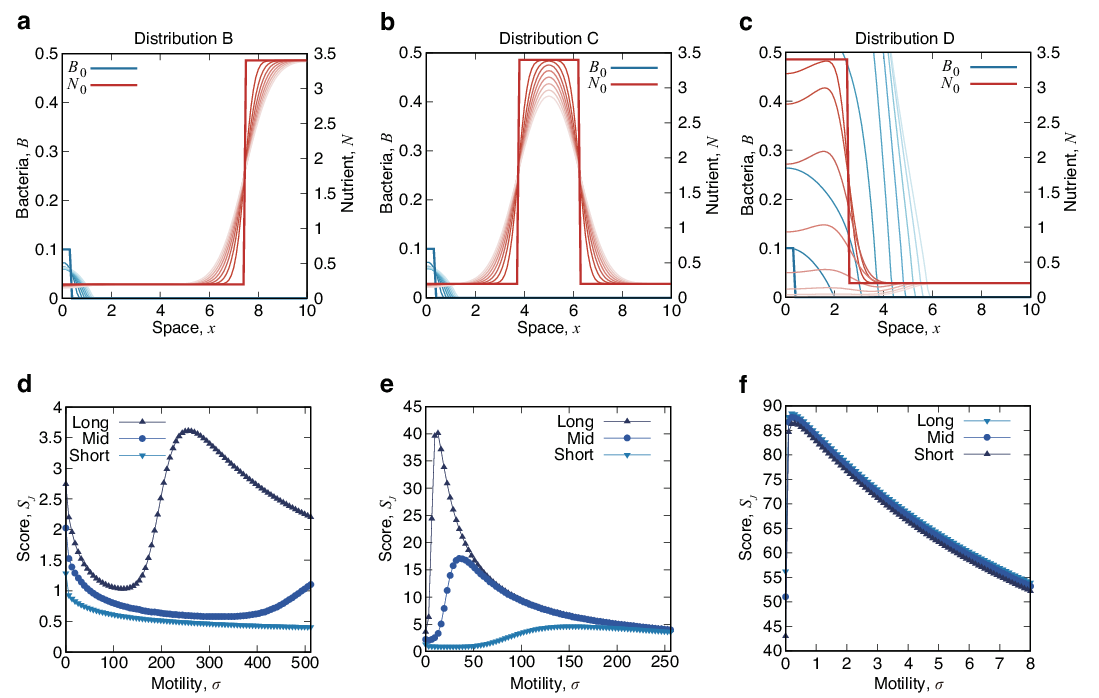}
\end{center}
\caption{Bacterial population model simulation of motility in linear response to nutrition for non-uniform nutrient distributions. 
{\bf a}, {\bf b}, {\bf c}, Initial distribution of bacteria and nutrients, $B_0 = B_0 (x)$, $N_0 = N_0 (x)$. The thin curves with color gradations are the time-series data from the early stages. In particular, the initial nutrient distribution $N_0 = N_0 (X)$ is non-uniform here (distribution B, C and D, respectively). 
{\bf d}, {\bf e}, {\bf f}, Dependence of growth strategy score $S_J$ on motility, where $J \in \left\{ {\rm Short},\  {\rm Mid}, \ {\rm Long} \right\}$. The initial nutrient distribution $N_0 = N_0 (X)$ is distributions B ({\bf d}), C ({\bf e}) and D ({\bf f}), respectively). }
\label{fig:2}
\end{figure*}

Furthermore, simulation results confirm such general principles under different simulation periods $J \in \left\{ {\rm Short},\  {\rm Mid}, \ {\rm Long} \right\}$ (Fig.~\ref{fig:1}c). The growth strategy score $S_J$ increases as $\sigma$ increases, and reduces gradually when motility is too intense compared to the optimal strategy. Although the difference was insignificant between short and mid period, how $\sigma$ achieving highest score changes under different time spans follows the expectation that short-term strategies favor high motility by having less influence on energy consumption.

Simulations are also conducted under different environmental conditions, where initial nutrients are distributed in a non-homogenous way (Fig.~\ref{fig:2}). 
The results suggest that though the linear model generally performs well when the nutrient is distributed evenly or close to the inoculation point~(Fig.~\ref{fig:1}, \ref{fig:2}a), it cannot provide access to distant nutrients when the area around the inoculation point lacks nutrients~(Fig.~\ref{fig:2}b, c).

\subsection{Non-monotonic Response to Nutrition}

\begin{figure*}[tbp]
\begin{center}
\includegraphics[width=90mm]{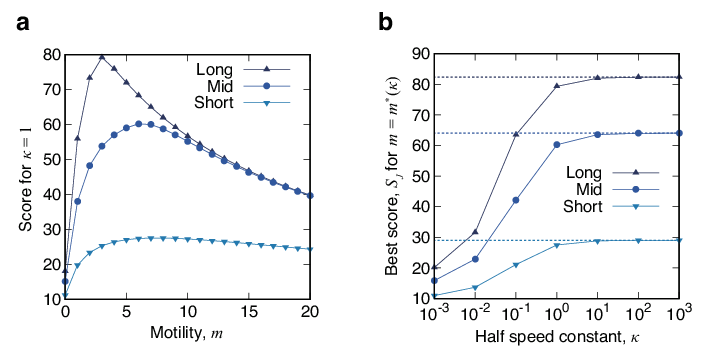}
\end{center}
\caption{Bacterial population model simulation of motility in non-monotonic response to nutrition for a uniform nutrient distribution (distribution A). 
The non-monotonic response is in the framework of a hyperbolic secant function. 
{\bf a}, Change in score for $m$ when $\kappa$ is fixed. There is one optimal peak $m = m_J^{\ast} ( \kappa )$ for a fixed $\kappa$, where $J \in \left\{ {\rm Short},\  {\rm Mid}, \ {\rm Long} \right\}$, respectively. 
{\bf b}, Dependence of maximum growth strategy score $S_J$ with $m = m_J^{\ast} ( \kappa )$ on $\kappa$, where $J \in \left\{ {\rm Short},\  {\rm Mid}, \ {\rm Long} \right\}$.
The dotted lines indicate the maximum scores in the linear response framework.}
\label{fig:3}
\end{figure*}

For each $\kappa$ under fixed conditions, there is only one peak $m = m^{\ast} (\kappa )$ of the highest $S_J$ as $m$ increases~(Fig.~\ref{fig:3}a). 
As $\kappa$ increases, the position of the peak $m = m^{\ast} (\kappa )$ also moves to the right on the $m$ axis.

Compared to the linear response, the inverted U-shaped response generally yields a similar score $S_J$ at homogeneous nutrient settings, each with best parameters applied. 
Here we use the best Score from linear response under each nutrient distribution in the previous section as a reference. 

Focusing on the best performance with corresponding parameters ($\sigma$ and $(\kappa,m)$), the simulation result of homogeneous media (nutrient distribution A) after $T_{\rm Short}$, $T_{\rm Mid}$, $T_{\rm Long}$, the highest $B_{\rm total}$ eventually converges to the highest $B_{\rm total}$ of linear variant simulation from below as $\kappa$ increases~(Fig.~\ref{fig:3}b).

\begin{figure*}[tbp]
\begin{center}
\includegraphics[width=135mm]{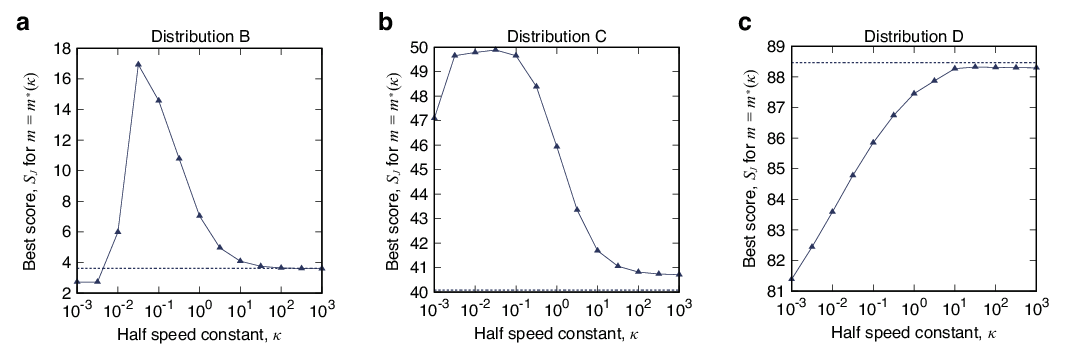}
\end{center}
\caption{Bacterial population model simulation of motility in non-monotonic response to nutrition for a non-uniform nutrient distribution. 
The non-monotonic response is in the framework of a hyperbolic secant function. 
The initial nutrient distributions $N_0 = N_0 (X)$ are corresponding to ({\bf a}) distribution B, ({\bf b}) distribution C, ({\bf c}) distribution D, while all results are recorded at $t = T_{\rm Long}$.
Each curve shows the dependence of  score $S_{\rm Long}$ with $m = m_{\rm Long}^{\ast} ( \kappa )$ on $\kappa$.
The $\kappa$ is sampled from $10^{-3}$ to $10^3$, in logarithmic sequence, and the $m$ that yields the maximum growth strategy score $S_{\rm Long}$ is selected for each corresponding $\kappa$.
The dotted lines indicate the maximum scores in the linear response framework as the reference value.}
\label{fig:4}
\end{figure*}

For inhomogeneous distribution, the simulation result demonstrates some of the noticeable behaviors from the non-linear motility-nutrient response that linear response is incapable to create.
Data suggests that on distribution B and C, where inoculation point is distant from the large portion of nutrient, the inverted U-shaped response achieves a higher score $S_J$ with $\kappa$ around 0.1, then gradually converges to the reference value from above~(Fig.~\ref{fig:4}a,b). 

The result from distribution D shows a similar tendency with homogeneous settings, as the $\kappa$ increases the best score $S_J$ converges to the reference value from below~(Fig.~\ref{fig:4}c).

By monitoring $N=N(x,t)$ and $B=B(x,t)$, we notice that the bacterial colony quickly passes through the low-concentration nutrient area and reaches the other side, acquiring extra territories and resources compared to its linear variants.

\section{Discussion}
Our investigation into model-based optimization reveals that bacterial motility strategies are governed by a delicate balance between immediate spatial expansion and long-term resource management. In the linear variant of our optimization problem, we observed a distinct divergence in population yield $B_{\rm total}$ curves based on the metabolic investment parameter, $\sigma$, under homogeneous nutrient conditions~(Fig.~\ref{fig:1}). These results strongly support the hypothesis that energy-saving strategies (lower $\sigma$) are advantageous over longer time horizons, whereas energy-consuming strategies (higher $\sigma$) prioritize short-term performance. However, we also identified a critical threshold where over-investment in motility becomes detrimental; excessively high $\sigma$ leads to a rapid depletion of resources, resulting in a significant decline in the overall welfare of the bacterial colony~(Fig.~\ref{fig:1}c). This suggests that for any given environmental setting and time period, there exists a singular optimal linear strategy that maximizes the population yield.

Under inhomogeneous conditions, the performance of these linear strategies is inherently sensitive to the spatial distribution of resources~(Fig.~\ref{fig:2}). While a moderate level of robustness is observed when the majority of nutrients remain accessible, the linear response lacks the flexibility required for more complex or unpredictable environments. This limitation is addressed by the non-linear (hyperbolic secant) motility response~(Fig.~\ref{fig:3}, \ref{fig:4}). Our findings indicate that with appropriate parameters ($\kappa$, $m$), non-linear responses can achieve performance comparable to linear ones in nutrient-abundant settings while avoiding the drawbacks of over-sensitivity. More importantly, when nutrients are located far from the initial inoculation point, this non-linear, non-monotonic response proves superior. It enables the colony to maintain a high level of robustness and ``searching ability,'' allowing for survival under starvation conditions by efficiently balancing exploration and nutrient consumption. 

These patterns suggest that sophisticated adaptive behaviors---which allow colonies to explore and colonize their environment effectively---can emerge from simple random motility regulated by local nutrient concentrations. Rather than relying on the ``intelligence'' of individual cells, the complex territory-expansion behaviors often attributed to directed chemotaxis can be interpreted as the functional efficacy of energy-aware random motility. This indicates that the regulatory mechanisms of random motility have been refined by environmental pressures to optimize collective outcomes without the need for complex sensing apparatus. While our current simulations did not provide sufficient data to isolate a specific energy-conserving response that suppresses motility at very high concentrations, we anticipate that such variants may emerge as optimal strategies within more extreme or fluctuating bacterial-nutrient dynamics.

In conclusion, our study demonstrates that the metabolic cost of motility is an essential component of bacterial survival strategies, particularly in resource-constrained or harsh environments. By explicitly incorporating these costs into a macroscopic PDE framework, we have quantified how environmental pressures shape motility preferences. The optimization results confirm that diverse environmental conditions favor varied motility strategies, providing a theoretical basis for the microbial diversity observed in nature. By maximizing population-level benefits like the population yield, these adaptive motility responses serve as a primary mechanism for surviving environmental selection. This framework not only clarifies the trade-offs between motility intensity and metabolic efficiency but also offers a powerful tool for predicting the collective dynamics of bacterial societies in complex ecosystems.

\backmatter


%

\bmhead{Acknowledgements}

This work was supported by JSPS KAKENHI, Grant Number 23K03208 (S.T.).





\section*{Declarations}

\bmhead{Conflict of interest} 

The authors have no conflict of interest to declare.



\bibliographystyle{sn-vancouver} 


\end{document}